\documentclass{an}
\usepackage{times}
\usepackage{graphicx}
\usepackage{fancyhdr}
\newcommand{\oversim}[2]{\protect{\mbox{\lower0.5ex\vbox{%
   \baselineskip=0pt\lineskip=0.2ex
   \ialign{$\mathsurround=0pt #1\hfil##\hfil$\crcr#2\crcr\sim\crcr}}}}} 
\begin{document}

\title{The distribution of ejected brown dwarfs in clusters}

\author{Simon\,P.\,Goodwin$^1$, David\,A.\,Hubber$^1$,
  Estelle\,Moraux$^2$ \and Anthony\,P.\,Whitworth$^1$}

\institute{$^1$School of Physics \& Astronomy, Cardiff University, 
5 The Parade, Cardiff, CF24 3YB, UK \\ $^2$ Insititute of Astronomy,
Madingley Road, Cambridge, CB}

\date{}

\abstract{
We examine the spatial distribution of brown dwarfs produced
  by the decay of small-$N$ stellar systems as expected from the
  embryo ejection scenario.  We model a cluster of several hundred
  stars grouped into 'cores' of a few stars/brown dwarfs.  These cores
  decay, preferentially ejecting their lowest-mass members.  Brown
  dwarfs are found to have a wider spatial distribution than stars,
  however once the effects of limited survey areas and unresolved
  binaries are taken into account it can be difficult to distinguish
  between clusters with many or no ejections.  A large
  difference between the distributions probably indicates that
  ejections have occurred, however similar distributions sometimes
  arise even with ejections.  Thus the spatial 
  distribution of brown dwarfs is not necessarily a good
  discriminator between ejection and non-ejection scenarios.
   }

\maketitle

\section{Introduction}

Brown dwarfs are observed to constitute some 15--25\% of the
objects in young star forming regions (e.g. Brice\~no et al. 2002; 
Muench et al. 2003; Luhman et al. 2003; Luhman 2005).  However, 
the formation mechanism(s) of brown dwarfs are currently unclear.  The
two most popular models are that (a) brown dwarfs form like stars but in
very low-mass cores (e.g. Greaves this volume; Padoan this volume), or (b)
brown dwarfs are stellar embryos that are ejected before than can
accrete sufficient mass to become stars (Reipurth \& Clarke 2001,2003).
(Also see Whitworth \& Goodwin (2005) and Whitworth \& Goodwin (this volume) 
for a review of other possible mechanisms).

Probably the most popular model for brown dwarf formation (at least
among theorists) is the ejection scenario (Reipurth \& Clarke 
2001).  This model appears to be supported by simulations of star 
formation in turbulent cores (e.g. Bate et al. 2002, 2003; Delgado 
Donate et al. 2004; Goodwin et al. 2004a,b).  

The ejection scenario has three significant problems.  Firstly, it
is not clear if ejected brown dwarfs are able to retain the 
significant discs which are observed around at least some young brown
dwarfs (e.g. Jayawardhana et al. 2003).  Secondly, whilst ejections 
do produce some brown dwarf-brown dwarf binary systems (e.g. Bate 
et al. 2002) a large population of brown dwarf-brown dwarf binaries 
would be difficult to explain via the ejection scenario (this problem
has recently become more acute with evidence that the brown
dwarf binary fraction may be significantly higher than previously
thought (Pinfield et al. 2003; Jeffries \& Maxted this volume).  Finally,
brown dwarfs are ejected with velocities of order 1 km s$^{-1}$ and it
may be expected that brown dwarfs are more widely distributed than
stars, in contrast to observations (Brice\~no et al. 2002; see also
Luhman 2005).

In this paper we concentrate on the final problem - the spatial
distribution of brown dwarfs.  In the ejection scenario, cores form
several objects in an unstable high-order multiple system.  Dynamical
interactions typically eject the lowest-mass members of the system until a
stable binary or hierarchical multiple remains.  We perform $N$-body 
simulations of a cluster of decaying small-$N$ `cores' to examine if
the distributions of brown dwarfs and stars are different or if the
velocity dispersion between cores effectively conceals the ejection of
brown dwarfs and low-mass stars.

\section{Method}

We simulate the $N$-body evolution of a cluster containing many
`cores' containing several stars using the NBODY6 code (Aarseth 
2003).  

A core is a small-$N$ system of stars and brown dwarfs in
which we assume accretion has finished and the gas reservoirs have been
exhausted {\em before} dynamical evolution occurs.  This is a
simplification to avoid dealing with the complex gas dynamics of the
combined accretion/ejection phase (e.g. Delgado Donate et al. 2004;
Goodwin et al. 2004a,b; see Umbreit et al. 2005 for an $N$-body treatment
of this problem).  However, we believe that it retains the key
physics of the current problem - the ejection of low-mass members of
cores which are moving relative to one-another within a cluster.

A cluster is initially composed of $N_{\rm core}$ cores each
containing $N_*$ stars and/or brown dwarfs (so that 
the total number of stars in the cluster is $N_{\rm tot} = 
N_{\rm core} \times N_*$).  For a typical distribution the inter-core
velocity dispersion is around $1$ km s$^{-1}$ (of order the ejection
velocities of low-mass stars and brown dwarfs - see below).

Within a core the masses of stars (for brevity `stars' generally 
means both stars {\em and} brown dwarfs) are randomly sampled 
from a Kroupa (2002) IMF of the form

\begin{displaymath}
N(M) \propto \left\{
\begin{array}{ll}
M^{-0.3} & \,\,\,\, 0.02 < M/M_\odot < 0.08 \\
M^{-1.3} & \,\,\,\, 0.08 < M/M_\odot < 0.5 \\ 
M^{-2.3} & \,\,\,\, 0.5 < M/M_\odot < 5 \\
\end{array} \right.
\end{displaymath}

\noindent The $N_*$ stars are distributed
randomly within a region of radius $R_{\rm core}$ and given random
velocities which are scaled such that the core is in virial
equilibrium.  We note that a random sampling of the IMF in this
way may not reproduce the correct (post-ejection) multiple system properties 
(Kroupa this volume).  

We choose $R_{\rm core} = 200$~au as the typical scale on which stars
are expected to form (e.g. Goodwin \& Kroupa 2005).  This is the scale 
at which a collapsing core
will reach the critical density at which the minimum mass for
fragmentation is reached ($\sim 10^{-13}$ g cm$^{-3}$).  We note 
that such a length scale has an observational
basis as the peak of the T~Tauri binary separation distribution occurs at
$\sim 100$~au (e.g. Mathieu 1994; Patience et al. 2002).

The cores so established are expected to decay on a timescale of 
$<0.1$~Myr rapidly if $N_{\rm star} > 2$ by ejecting the
lowest-mass members of the system until a stable hierarchical system or
a binary is formed (Anosova 1986; see also Sterzik \& Durisen 
2003; Hubber \& Whitworth 2005; Goodwin \& Kroupa 2005; Umbreit et
al. 2005).  We ignore the interaction of the protostars and the
ambient gas, as all we are interested in is the post-ejection velocity
distribution and we find that the ejection velocities of low-mass
stars and brown dwarfs are of order $1$ km s$^{-1}$, similar to that
found in more detailed simulations including gas (e.g. Umbreit et
al. 2005). 

Cores are then placed within the star cluster by positioning them
within a virialised Plummer sphere with a virial radius of 1~pc 
following the prescription of Aarseth et al. (1974).  

\section{Results}

We analyse the relative distributions of brown dwarfs and stars by
comparing the distances to the nearest neighbours (following
Brince\~no et al. 2003); using the mean distances to the 
nearest neighbour (nearest neighbour distance, or NND).

There are two important observational biases that must be included
when analysing the data.  The first is unresolved binaries, which we
include by ignoring the secondary component of any system if it is 
within 250~au of the primary (roughly 2 arcsec at the distance of
Taurus).  The second is that surveys are over limited areas: 
cluster members at great distances from the cluster centre will
normally not be found in surveys, and - even if they are - would be
difficult to unambiguously relate to the cluster without additional 
proper motion 
studies.  Therefore we restrict ourselves to stars and brown dwarfs 
within a projected distance of 5~pc from the cluster centre (this
corresponds to an area of 13 square degrees at the distance of
Taurus).

\begin{figure*}[h]
{\includegraphics[width=6cm,angle=270]{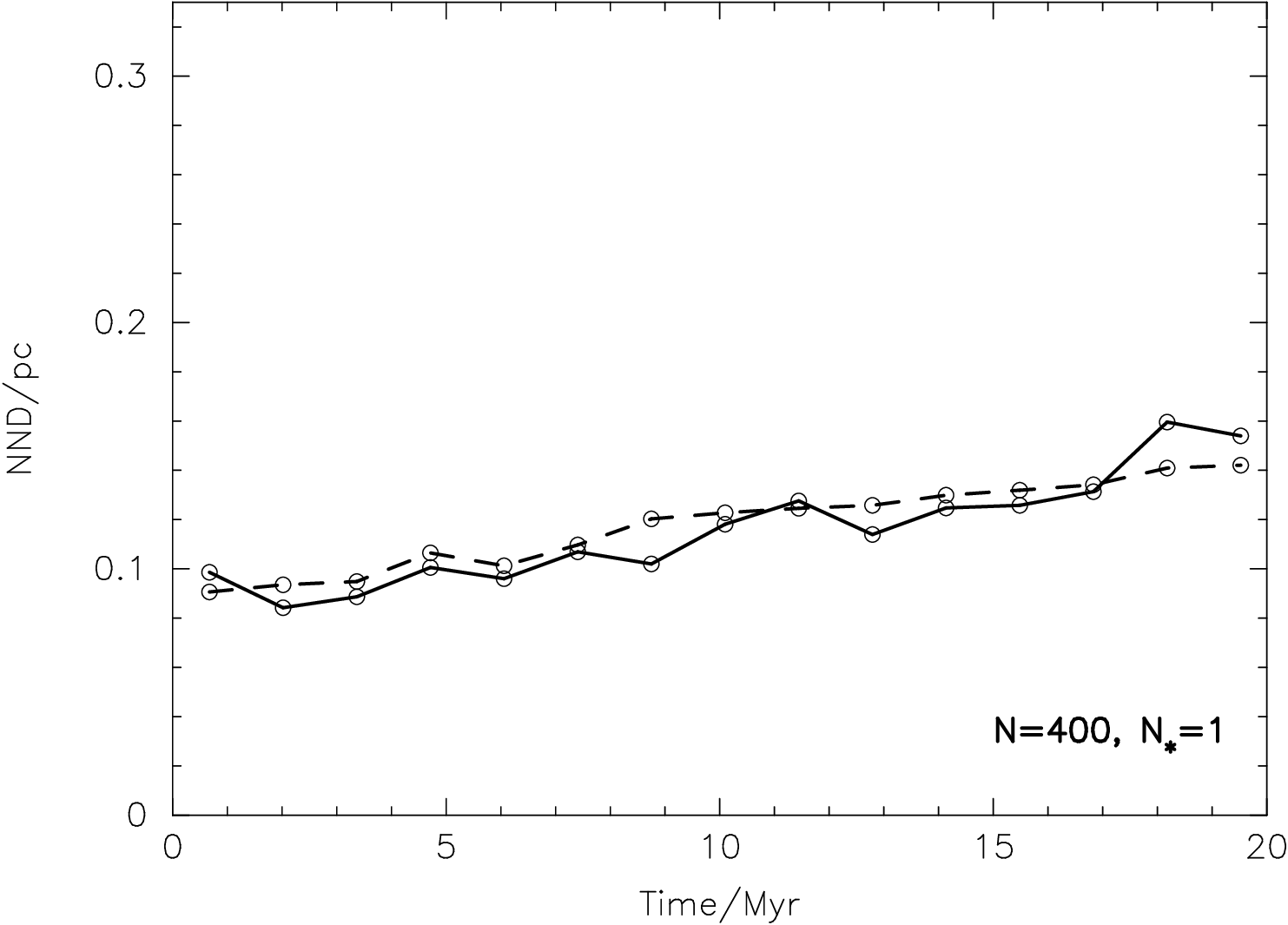}} 
{\includegraphics[width=6cm,angle=270]{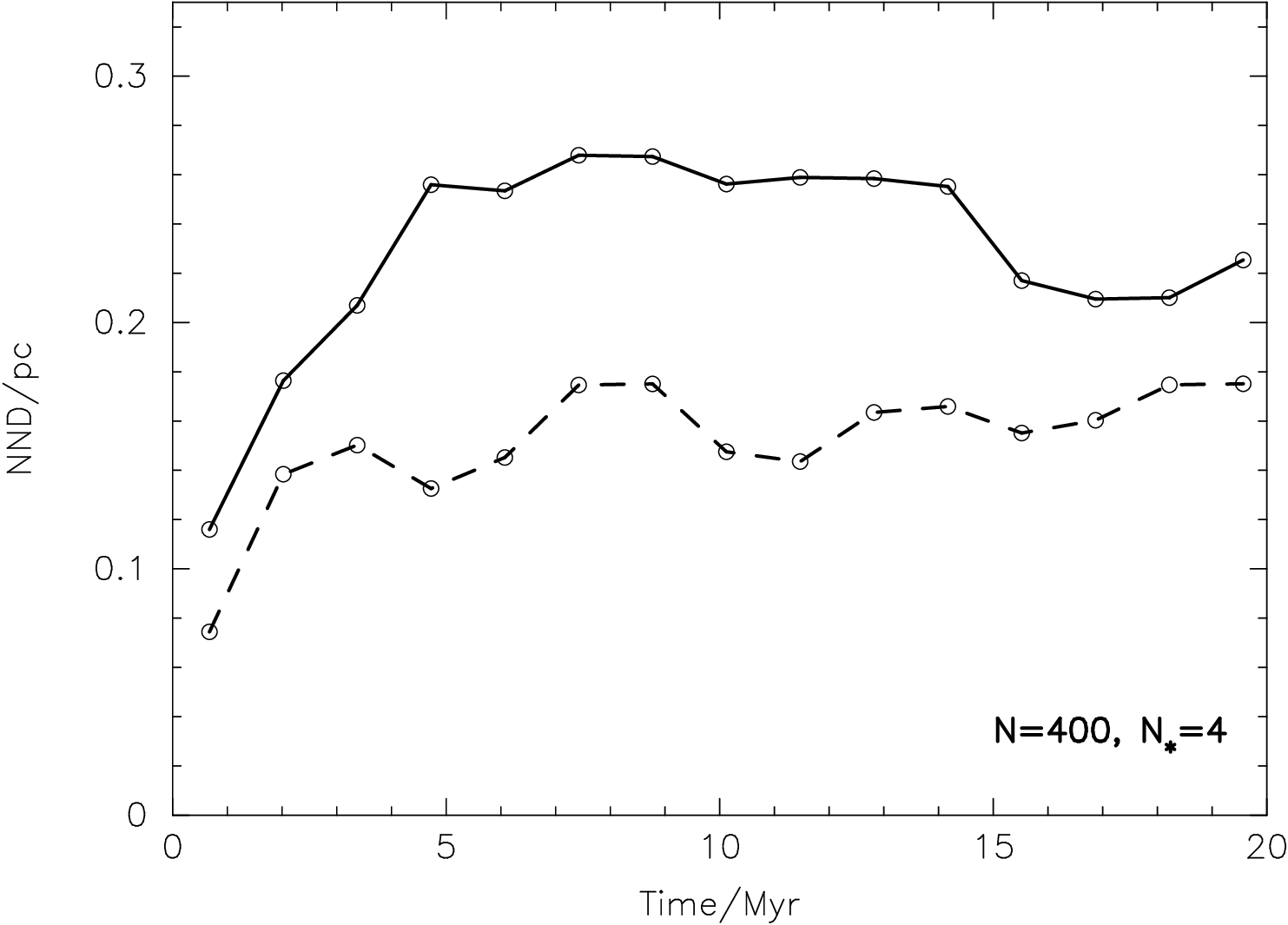}}
\caption{The average nearest neighbour distances (NND) of stars
  (dahsed-line) and brown dwarfs (full line) for clusters with $N_{\rm
  tot}=400$ with $N_*=1$ (left panel) and $N_*=4$ (right panel).}
\end{figure*}

In Fig.~1 we compare the stellar and brown dwarf NNDs for clusters
with a total of $N_{\rm tot} = 400$ stars and $N_* = 1$ (ie. no decay
of groups within cores) and $N_* = 4$ (generally decay into a binary
and two single stars).  

When $N_* = 1$, the NNDs of the stars and
brown dwarfs are indistinguishable, over the 20~Myr of cluster
evolution that is followed, the NNDs increase somewhat as the cluster
expands slightly through 2-body interactions.  Over a significant
time, we would expect the brown dwarf NND to become larger than the
stellar NND as low-mass stars and brown dwarfs gain a higher velocity
dispersion as equipartition is established through 2-body encounters
(however, in reality the cluster will probably disperse long before
this becomes important).

In contrast, when $N_* = 4$, the effect of the very rapid decay of the
small-$N$ cores is to produce a population of more widely dispersed
brown dwarfs and low-mass stars with a larger NND.  (It should be
noted that not accounting for unresolved binaries and limited survey
areas makes these differences significantly more extreme).  Towards
the end of the simulation the NND for brown dwarfs is seen to drop
significantly.  This is caused by a number of brown dwarfs escaping
from the cluster at late times, either because they were ejected with
an initially low velocity or have gained velocity due to later encounters in 
the cluster.

\begin{figure*}
{\includegraphics[width=6cm,angle=270]{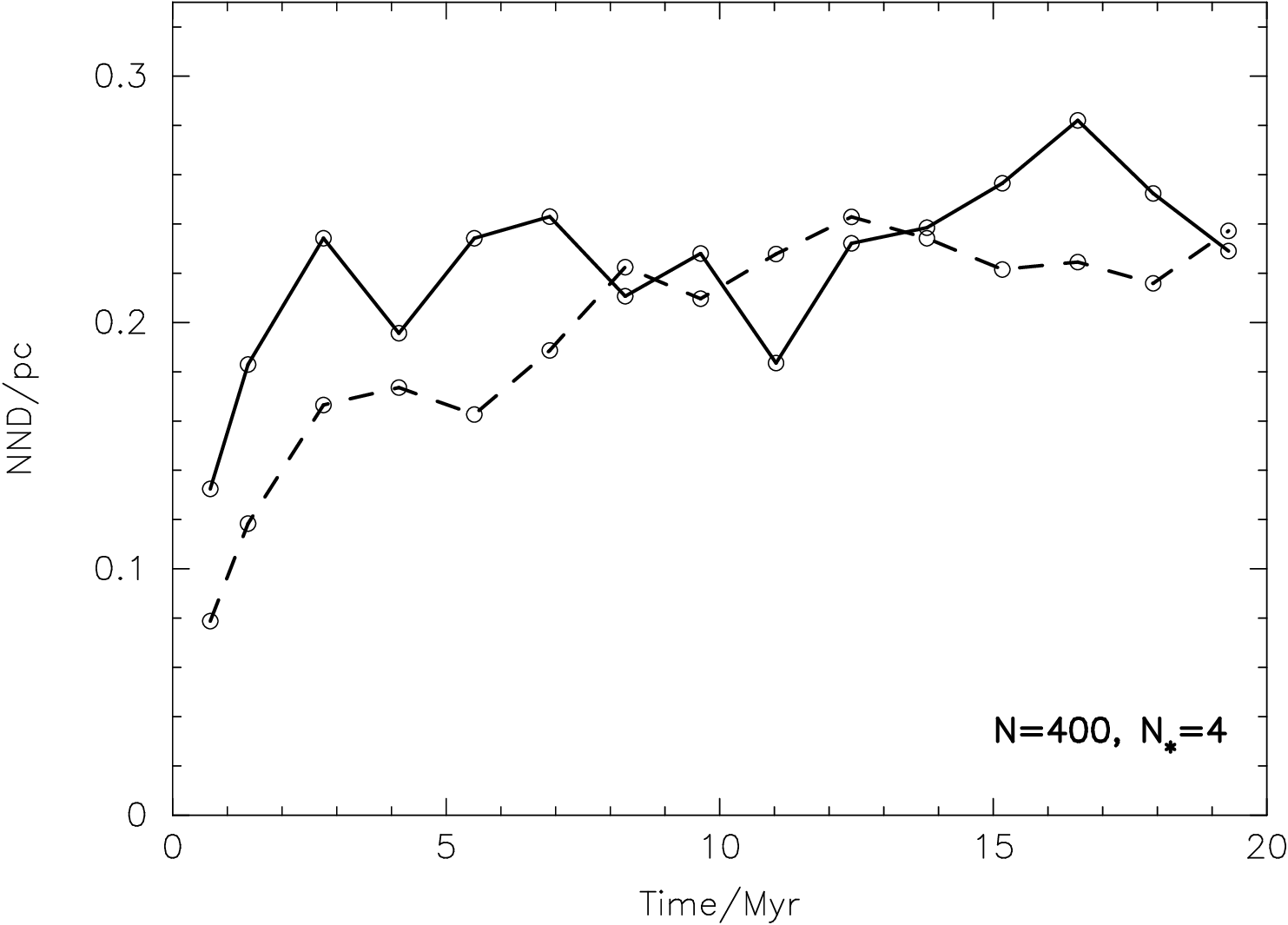}} 
{\includegraphics[width=6cm,angle=270]{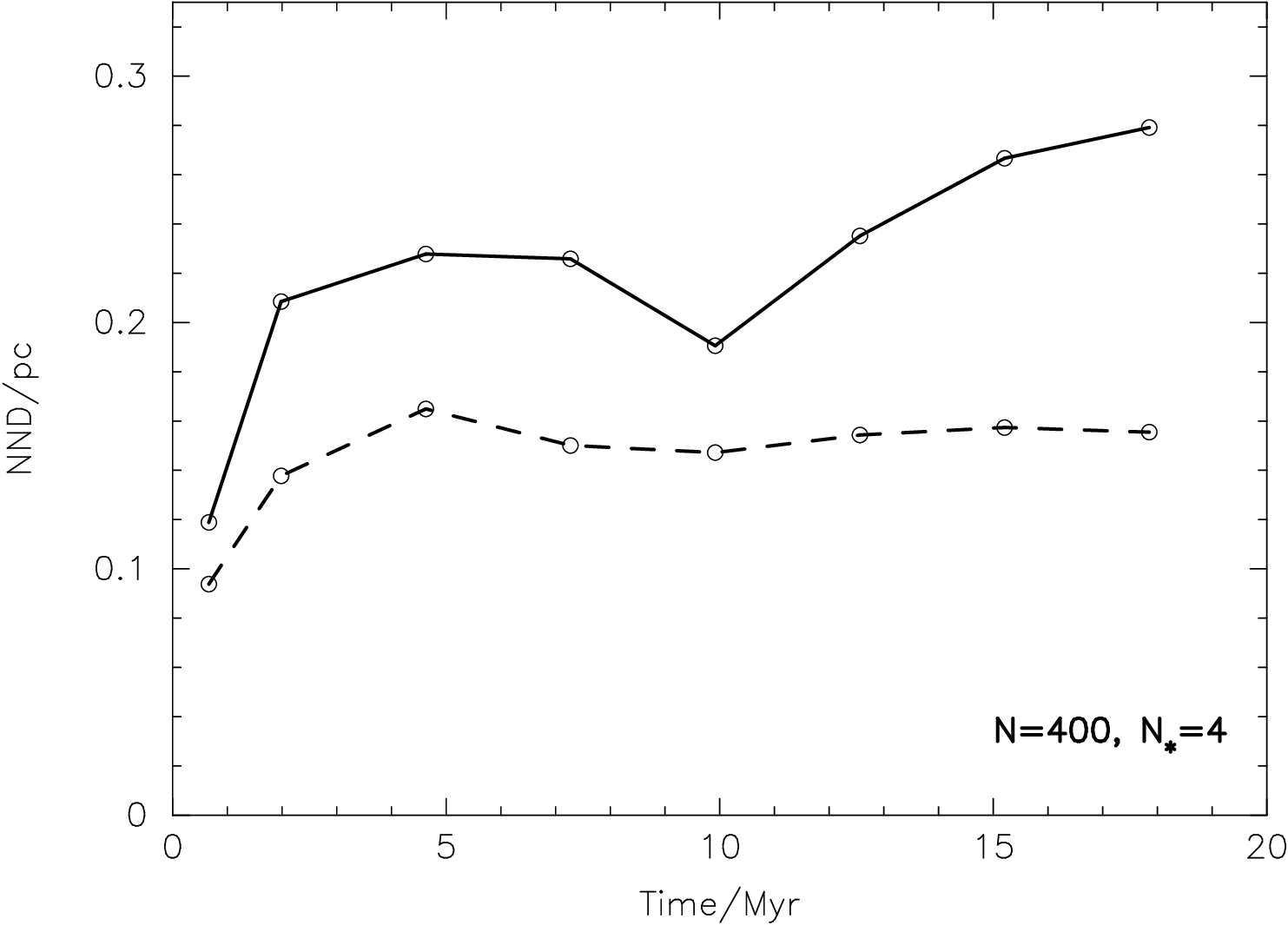}} \\
{\includegraphics[width=6cm,angle=270]{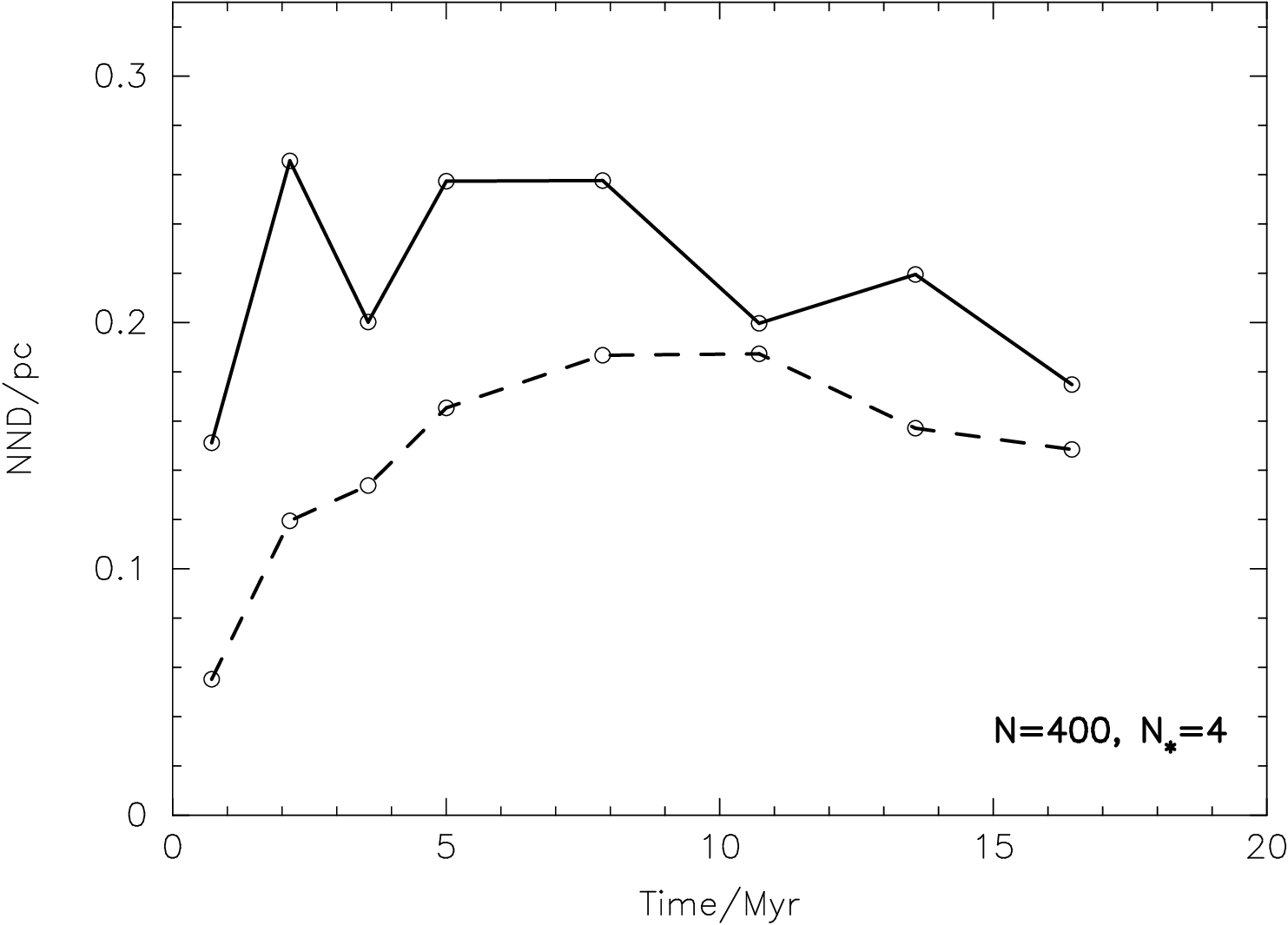}} 
{\includegraphics[width=6cm,angle=270]{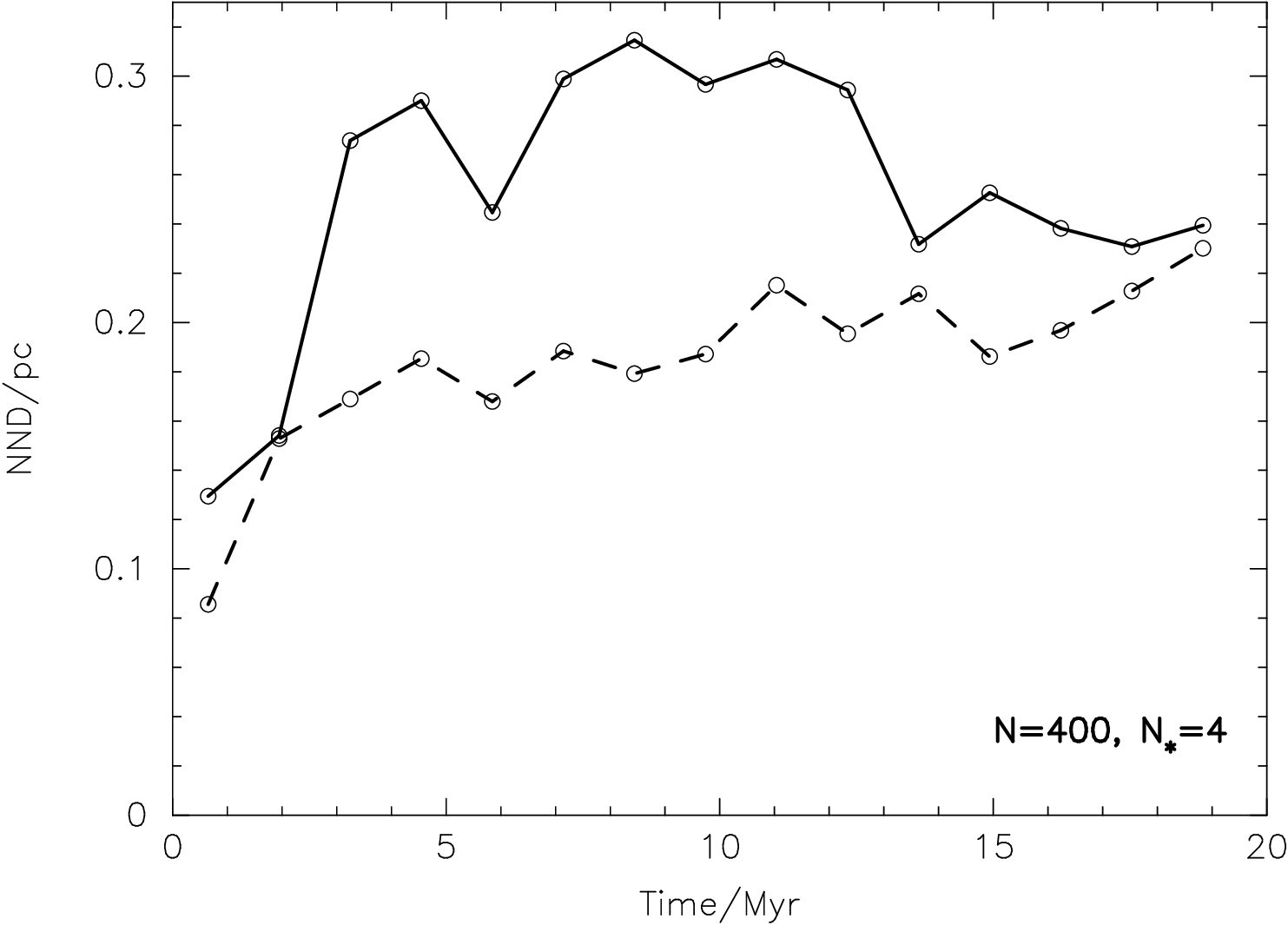}} 
\caption{The average nearest neighbour distances (NND) of stars
  (dashed-line) and brown dwarfs (full line) for clusters with $N_{\rm
  tot}=400$ and $N_*=4$ for four different realisations (different
  random number seeds).}
\end{figure*}

Thus it would appear that the ejection of brown dwarfs from small-$N$
cores can produce a significant and observable effect in the spatial
distributions.  However, ejections do not always produce a significant
difference in the spatial distributions.  In Fig.~2 we show four 
more simulations with $N_*=4$ (making 5 in
total including the simulation from Fig.~1).  The only difference
between these simulations is the random number seed used to generate
the initial conditions.  In the final simulation in particular there
is no significant difference between brown dwarfs and stars at any
time.  

The most extreme differences between the
NNDs of brown dwarfs and stars occurs at 5 - 10~Myr: after ejections
have had time to significantly disperse the brown dwarfs, but before
most of them have escaped the inner regions of the cluster (travelling
at 1~km s$^{-1}$ this should take $\sim 5$~Myr).  However, most young
clusters are observed at ages of $\sim 1$~Myr (e.g. the Taurus
observations of Brice\~no et al. 2002) at which point only 1 of the 5
simulations shows a very significant difference.

\section{Conclusions}

When ejections are not important ($N_* = 1$) the spatial
distributions of stars and brown dwarfs are (unsurprisingly) very
similar.  When ejections become important ($N_*=4$) then the spatial
distributions can show significant differences.  However, these
differences can disappear altogether in some clusters depending on the
exact details of the initial clustering.  In addition, at the young age
of most well-studied young clusters a significant difference in the
spacial distributions is seen in only 1 out of 5 simulations.

Thus if brown dwarfs and stars have different spatial distributions it
is probably a signature of ejections, however the lack of a difference
does not necessarily exclude the ejection scenario.  

We suggest that binarity is a far stronger discriminator between 
models since (as yet) there is no way in which to make significant 
numbers of brown dwarf-brown dwarf binaries from ejections.  However,
we are currently investigating if a more sophisticated statistical 
analysis of positions may yield a more robust discriminator.

The discovery of many more brown dwarfs in Taurus extending over a 
wide area (Guieu et al. this volume) does suggest that ejections may 
have been responsible for at least some of the Taurus brown dwarfs.

\acknowledgements

SPG is supported by a UKAFF Fellowship.  DAH acknowledges the support
of a PPARC studentship.

\end{document}